\documentclass[10pt,twocolumn,twoside]{IEEEtran}
\pdfoutput=1
\usepackage{xcolor}
\usepackage{graphicx}
\usepackage{tikz}
\usepackage{color}
\usepackage{amsmath}
\usepackage{subfig}
\usepackage{amssymb}
\usepackage{float}
\graphicspath{{./Fig/}}
\usepackage{math}

\usepackage[final]{review}

\title{
Estimation of the Direct-Path Relative Transfer Function for Supervised Sound-Source Localization
}

\author{Xiaofei Li,
        Laurent~Girin,
        Radu Horaud and
        Sharon Gannot
\thanks{X. Li and R. Horaud are with INRIA Grenoble Rh\^one-Alpes, Montbonnot Saint-Martin, France. 
E-mail: \texttt{first.last@inria.fr}
}
\thanks{L. Girin is with INRIA Grenoble Rh\^one-Alpes and with Univ. Grenoble Alpes, GIPSA-lab, Grenoble, France. 
E-mail: \texttt{laurent.girin@gipsa-lab.grenoble-inp.fr}
}%
\thanks{Sharon Gannot is with Bar Ilan University, Faculty of Engineering, Israel. 
E-mail: \texttt{Sharon.Gannot@biu.ac.il}
}
\thanks{X. Li, L. Girin and R. Horaud acknowledge support from the EU FP7 STREP project EARS \#609465 and from 
the ERC Advanced Grant VHIA \#340113.}
}
\begin{document}
\bibliographystyle{ieeetr}

\maketitle
                            
\begin{abstract}
This paper addresses the problem of binaural localization of a single speech source in noisy and reverberant environments. 
\addnote[cla_abs]{1}{For a given binaural microphone setup, the binaural response corresponding to the direct-path propagation of a single source is a function of the source direction. } 
In practice, this response is contaminated by noise and reverberations. The direct-path relative transfer function (DP-RTF) is defined as the ratio between the direct-path acoustic transfer function of the two channels. We propose a method to estimate the DP-RTF from the noisy and reverberant microphone signals in the short-time Fourier transform domain.
First, the convolutive transfer function approximation is adopted to accurately represent the impulse response of the sensors in the STFT domain. Second, the DP-RTF is estimated by using the auto- and cross-power spectral densities at each frequency and over multiple frames. 
In the presence of stationary noise, an inter-frame spectral subtraction algorithm is proposed, which enables to achieve the estimation of noise-free auto- and cross-power spectral densities. Finally, the estimated DP-RTFs are concatenated across frequencies and used as a feature vector for the localization of speech source. Experiments with both simulated and real data show that the proposed localization method performs well, even under severe adverse acoustic conditions, and outperforms state-of-the-art localization methods under most of the acoustic conditions. 
\end{abstract}


\section{Introduction}
\label{sec:introduction}
Sound-source localization (SSL) is an important task for many applications, e.g., robot audition, video conferencing, hearing aids, to cite just a few.  
In the framework of human-inspired binaural hearing, two interaural cues are widely used for SSL, namely the interaural phase difference (IPD) and the interaural level difference (ILD) \cite{viste2004,willert2006,stern2006,raspaud2010,woodruff2012,deleforge2015acoustic,deleforge2015colocalization}. \addnote[cla_general]{1}{In the general case where the sensor array is not free-field, i.e. the microphones are placed inside the ears of a dummy head or on a robot head, the interaural cues are frequency-dependent due to the effects on sound propagation induced by the shape of the outer ears, head and torso \cite{blauert1997}. This is true even for anechoic recordings, i.e. in the absence of reverberations. SSL is then based on the relationship between interaural cues and direction of arrival (DOA) of the emitting source. }

When the short-time Fourier transform (STFT) is used, the ILD and IPD correspond to the magnitude and argument, respectively, of the relative transfer function (RTF), which is the ratio between the acoustic transfer functions (ATF) of the two channels \cite{gannot2001}.
In a reverberant environment, the RTF contains both direct-path information, namely the direct wave propagation path from the source location to the microphone locations, and information representing early and late reverberations. Extracting the direct path is of crucial importance for SSL. In an anechoic and noise-free environment the source direction can be easily estimated from the RTF. However, in practice, noise and reverberations are often present and contaminate SSL estimation.


In the presence of noise, based on the stationarity of the noise and the non-stationarity of the desired signal, the RTF was estimated in \cite{gannot2001} by solving a set of linear equations, and in \cite{dvorkind2005} by solving a set of nonlinear decorrelation equations. In \cite{dvorkind2005}, the time difference of arrival (TDOA) was estimated based on RTF, and a TDOA tracking method was also proposed. 
These methods have the limitation that a significant amount of noisy frames are included in the estimation. An RTF identification method based on the probability of speech presence and on spectral subtraction was proposed in \cite{cohen2004}: this method uses only the frames which are highly likely to contain speech. 
The unbiased RTF estimator proposed in \cite{mine2015assp} is based on segmental power spectral density matrix subtraction, which is a more efficient method to remove noise compared with the approaches just mentioned.
The performance of these spectral subtraction techniques was analyzed and compared with eigenvalues decomposition techniques in \cite{markovich2015}.

The RTF estimators mentioned above assume a multiplicative transfer function (MTF) approximation \cite{avargel2007spl}, i.e., the source-to-microphone filtering process is assumed to be represented by a multiplicative process in the STFT domain. Unfortunately, this is only justified when the length of the filter impulse response is shorter than the length of the STFT window, which is rarely the case in practice. 
Moreover, the RTF is usually estimated from the ratio between two ATFs that include reverberation, rather than from the ratio between ATFs that only correspond to the direct-path sound propagation. Therefore, currently available RTF estimators are poorly suitable for SSL in reverberant environments. 

The influence of reverberation on the interaural cues is analyzed in \cite{zannini2011}. The relative early transfer function was introduced in \cite{schwartz2015} to suppress reverberation. 
Several techniques were proposed to extract the RTF that corresponds to the direct-path sound propagation, e.g., based on detecting time frames with less reverberations. The precedence effect, e.g., \cite{litovsky1999}, widely used for SSL, relies on the principle that signal onsets are dominated by the direct path. 
Based on band-pass filter banks, the localization cues are extracted only from reliable frames, such as the onset frames in \cite{bechler2005}, 
the frames preceding a notable maximum \cite{heckmann2006}, the frames weighted by the precedence model \cite{hummersone2013}, etc. Interaural coherence was proposed in \cite{faller2004} to select binaural cues not contaminated by reverberations.
Based on Fourier transform, the coherence test \cite{mohan2008}, and the direct-path dominance test \cite{nadiri2014} are proposed to detect the frames dominated by one active source, from which  localization cues can be estimated.
However, in practice, there are always reflection components in the frames selected by these methods, due to an inaccurate model or an improper decision threshold. 

\textbf{Contributions and Method Overview:} In this paper, we propose a direct-path RTF estimator suitable for the localization of a single speech-source in noisy and reverberant environments. 
We build on the cross-band filter proposed in \cite{avargel2007} for system identification in the STFT domain. 
This filter represents the impulse response in the STFT domain by a cross-band convolutive transfer function instead of the multiplicative (MTF) approximation.
In practice we consider the use of a simplified convolutive transfer function (CTF) approximation, as used in \cite{talmon2009}. 
The first coefficient of the CTF at different frequencies represents the STFT of the first segment of the channel impulse response, 
which is composed of the direct-path impulse response, plus possibly few early reflections. 
In particular, if the time delay between the direct-path wave and the first notable reflection is large, less reflections are included. 
Therefore, we refer to the first coefficient of the CTF as the direct-path acoustic transfer function, and the ratio between the coefficients from two channels is referred to as the \textit{direct-path relative transfer function} (DP-RTF).

Inspired by \cite{benesty2000} and based on the relationship of the CTFs between the two channels, we use the auto- and cross-power spectral densities (PSD) estimated over multiple STFT frames,  to construct a set of linear equations in which the DP-RTF is the unknown variable. 
Therefore, the DP-RTF can be estimated via standard least squares. 
In the presence of noise, an inter-frame spectral subtraction technique is proposed, extending our previous work \cite{mine2015assp}. The auto- and cross-PSD estimated in a frame with low speech power are subtracted from the PSDs estimated in a frame with high speech power. 
After subtraction, low noise power and high speech power are left due to the stationarity of the noise and the non-stationarity of the speech signal. The DP-RTF is estimated using the remaining signal's auto- and cross-PSD.
This PSD subtraction process does not require an explicit estimation of the noise PSD, hence it does not suffer from noise PSD estimation errors.

Finally, the estimated DP-RTFs are concatenated over frequencies and plugged into an SSL method, e.g., \cite{deleforge2015acoustic}. Experiments with simulated and real data were conducted under various acoustic conditions, e.g., different reverberation times, source-to-sensor distances, and signal-to-noise ratios. The experimental results show that the proposed method performs well, even in adverse acoustic conditions, and outperforms the MTF-based method \cite{mine2015assp}, the coherence test method \cite{mohan2008} and the conventional SRP-PHAT method in most of the tested conditions. 

The remainder of this paper is organized as follows. Section~\ref{sec:ctf} formulates the sensor signals based on the crossband filter. Section~\ref{sec:dprtf} presents the DP-RTF estimator in a noise-free environment.  
The DP-RTF estimator in the presence of noise is presented in Section~\ref{sec:dprtfn}. In Section~\ref{sec:ssl}, the SSL algorithm is described.  Experimental results are presented in Section~\ref{sec:experiments1} and \ref{sec:experiments2}, and Section~\ref{sec:conclusion} draws some conclusions.

\section{Cross-band Filter and Convolutive Transfer Function}
\label{sec:ctf}
We consider first a non-stationary source signal $s(n)$, e.g., speech, emitted in a noise-free environment. The received binaural signals are
\begin{align}\label{xn}
\begin{array}{l}
 x(n)=s(n)\star a(n)\\
 y(n)=s(n)\star b(n), 
 \end{array}
\end{align}
where $\star $ denotes convolution, and $a(n)$ and $b(n)$ are the binaural room impulse responses (BRIR) from the source to the two microphones. \addnote[exp_BRIR]{1}{The BRIRs combine the effects of the room acoustics (reverberations) and the effects of the sensor set-up (e.g., dummy head/ears). }
Applying the STFT, (\ref{xn}) is approximated in the time-frequency (TF) domain as
\begin{align}\label{xpk}
\begin{array}{l}
 x_{p,k}=s_{p,k}\; a_k \\
 y_{p,k}=s_{p,k}\; b_k,
 \end{array}
\end{align}
where $x_{p,k}$, $y_{p,k}$ and $s_{p,k}$ are the STFT of the corresponding signals ($p$ is the time frame index and $k$ is the frequency bin index), and $a_k$ and $b_k$ are the ATFs corresponding to the BRIRs. Let $N$ denote the length of a time frame or, equivalently, the size of the STFT window. 
Eq.~(\ref{xpk}) corresponds to the MTF approximation, which is only valid when the impulse response $a(n)$ is shorter than the STFT window. 
In the case of non-stationary acoustic signals, such as speech, a relatively small value for $N$ is typically chosen to assume \textit{local} stationarity, i.e., within a frame. Therefore, the MTF approximation (\ref{xpk}) is questionable in a reverberant environment, since the room impulse response could be much longer than the STFT window. 

To address this problem cross-band filters were introduced  \cite{avargel2007} to represent more accurately a linear system with long impulse response in the STFT domain. 
Let $L$ denote the frame step.  
The cross-band filter model consists in representing the STFT coefficient $x_{p,k}$ in (\ref{xpk}) as a summation over multiple convolutions across frequency bins (there is an equivalent expression for $y_{p,k}$):
\begin{align}\label{xpk2}
 x_{p,k} = \sum_{p'=-C}^{Q_k-1} \sum_{k'=0}^{N-1} s_{p-p',k'} \; a_{p',k',k}.
\end{align}
From \cite{avargel2007}, if $L<N$, then $a_{p',k',k}$ is non-causal, with $C = \lceil N/L \rceil -1$ non-causal coefficients. 
The number of causal filter coefficients  $Q_k$ is related to the reverberation time at the $k$-th frequency bin, which will be discussed in detail in Section~\ref{sec:experiments1}.
The TF-domain impulse response $a_{p',k',k}$ is related to the time-domain impulse response $a(n)$ by:
\begin{align}\label{hp}
a_{p',k',k}={(a(n)\star \zeta_{k,k'}(n))}|_{n=p'L},
\end{align}
which represents the convolution with respect to the time index $n$ evaluated at frame steps, with
\begin{align}\label{phik}
\zeta_{k,k'}(n) = e^{j\frac{2\pi}{N}k'n}\sum_{m=-\infty}^{+\infty} \overline{\omega}(m) \: \omega(n+m) \: e^{-j\frac{2\pi}{N}m(k-k')},
\end{align}
where $\overline{\omega}(n)$ and $\omega(n)$ denote the STFT analysis and synthesis windows, respectively.
A convolutive transfer function (CTF) approximation is further introduced and used in \cite{talmon2009} to simplify the analysis, i.e., only band-to-band filters are considered, $k=k'$. 
Hence, (\ref{xpk2}) is rewritten as
\begin{align}
\label{eq:xpk3}
 x_{p,k} = \sum_{p'=0}^{Q_k-1} s_{p-p',k}a_{p',k}= s_{p,k}\star a_{p,k},  
 \end{align}
where we assumed $L\approx N$ such that non-causal coefficients are disregarded. Note that $a_{p',k',k}$ is replaced with $a_{p',k}$ to simplify the notations.
The cross-band filter and CTF formalism will now be used to extract the impulse response of the direct-path propagation.

\section{Direct-Path Relative Transfer Function}
\label{sec:dprtf}
From (\ref{hp}) and (\ref{phik}), with $k'=k$ and $p'=0$, the first coefficient of $a_{p',k}$ in the CTF approximation \eqref{eq:xpk3} can be derived as
\begin{align}
 a_{0,k} = ({a(n)\star \zeta_{k,k}(n)})|_{n=0} 
& =\sum_{t=0}^{T-1} a(t)\zeta_{k,k}(-t) \nonumber \\
 &=\sum_{t=0}^{N-1} a(t)\nu(t)e^{-j\frac{2\pi}{N}kt},
\end{align}
where  $T$ is the length of the BRIR and
\begin{equation}
 \nu(n)=
 \begin{cases} \sum_{m=0}^{N}\overline{\omega}(m)\omega(m-n)  & \mbox{if } 1-N\le n\le N-1, \\
  0, & \mbox{otherwise.}
 \end{cases} \nonumber
\end{equation}
Therefore, $a_{0,k}$ (as well as $b_{0,k}$) can be interpreted as the $k$-th Fourier coefficient of the impulse response segment $a(n)|_{n=0}^{N-1}$ windowed by $\nu(n)|_{n=0}^{N-1}$.
Without loss of generality, we assume that the room impulse responses $a(n)$ and $b(n)$ begin with the impulse responses of the direct-path propagation.    
If the frame length $N$ is properly chosen, $a(n)|_{n=0}^{N-1}$ and $b(n)|_{n=0}^{N-1}$ are composed of the impulse responses of the direct-path and a few reflections. 
Particularly, if the initial time delay gap (ITDG), i.e. the time delay between the direct-path wave and the first notable reflection, is large compared to $N$, $a(n)|_{n=0}^{N-1}$ and $b(n)|_{n=0}^{N-1}$ mainly contain the direct-path impulse response. Therefore we refer to $a_{0,k}$ and $b_{0,k}$ as the direct-path ATFs. 
By definition, the DP-RTF is given by (we remind that the direct path is relevant for sound source localization):
\begin{align}\label{eq:dp-rtf}
 d_{k} = \frac{b_{0,k}}{a_{0,k}}.
\end{align}
\addnote[exp_DP]{1}{In summary, the CTF approximation offers a nice framework to encode the direct-path part of a room impulse response into the first CTF coefficients. Applying this to each channel of a BRIR and taking the ratio between the first CTF coefficients of each channel provides the DP-RTF. Of course, in practice, the DP-RTF must be estimated from the sensor signals. }

\subsection{Direct-Path Estimation}
\label{sec:dprtf:estimation}

Since both channels are assumed to follow the CTF model, we can write:
\begin{align}\label{xyha}
 x_{p,k}\star b_{p,k}=s_{p,k}\star a_{p,k}\star b_{p,k}=y_{p,k}\star a_{p,k}.
\end{align}
This relation was proposed in \cite{benesty2000,benesty1995} for the time-domain TDOA estimation and is here extended to the CTF domain.
In vector form \eqref{xyha} can be written as
\begin{align}\label{mxa}
 \mathbf{x}_{p,k}\tp \mathbf{b}_k = \mathbf{y}_{p,k}\tp \mathbf{a}_k,
\end{align}
where $\tp$ denotes vector or matrix transpose, and 
\begin{align}
 \mathbf{x}_{p,k} &= [x_{p,k},x_{p-1,k},\dots,x_{p-Q_k+1,k}]\tp, \nonumber \\
 \mathbf{y}_{p,k} &= [y_{p,k},y_{p-1,k},\dots,y_{p-Q_k+1,k}]\tp, \nonumber \\
 \mathbf{b}_k &= [b_{0,k},b_{1,k},\dots,b_{Q_k-1,k}]\tp, \nonumber \\
 \mathbf{a}_k &= [a_{0,k},a_{1,k},\dots,a_{Q_k-1,k}]\tp. \nonumber
\end{align}
Dividing both sides of (\ref{mxa}) by $a_{0,k}$ and reorganizing the terms, we can write:
 \begin{align}\label{zpk}
 y_{p,k} = \mathbf{z}_{p,k}\tp  \mathbf{g}_k,
 \end{align}
where
\begin{align}
 \mathbf{z}_{p,k} &=[x_{p,k},\dots,x_{p-Q_k+1,k},y_{p-1,k},\dots,y_{p-Q_k+1,k}]\tp, \nonumber \\
 \mathbf{g}_k &=\left[\frac{b_{0,k}}{a_{0,k}},\dots,\frac{b_{Q_k-1,k}}{a_{0,k}},-\frac{a_{1,k}}{a_{0,k}},\dots,-\frac{a_{Q_k-1,k}}{a_{0,k}}\right]\tp. \nonumber
\end{align}
We see that the DP-RTF appears as the first entry of $\mathbf{g}_k$. Hence, in the following, we base the estimation of the DP-RTF on the construction of $y_{p,k}$ and $\mathbf{z}_{p,k}$ statistics.
More specifically, multiplying both sides of (\ref{zpk}) by $y_{p,k}^*$ (the complex conjugate of $y_{p,k}$) and taking the expectation, $E\{\cdot\}$, we obtain:
\begin{align}\label{phi}
\phi_{yy}(p,k) = \phivect_{zy}\tp(p,k) \: \mathbf{g}_k,
\end{align}
where $\phi_{yy}(p,k)=E\{y_{p,k}y_{p,k}^{*}\}$ is the PSD of $y(n)$ at TF bin $(p,k)$, and
\begin{align}
 \phivect_{zy}(p,k) = & [E\{x_{p,k}y_{p,k}^*\},\dots,E\{x_{p-Q_k+1,k}y_{p,k}^*\}, \nonumber \\
 &E\{y_{p-1,k}y_{p,k}^*\},\dots,E\{y_{p-Q_k+1,k}y_{p,k}^*\}]\tp \nonumber
\end{align}
is a vector composed of cross-PSD terms between the elements of $\mathbf{z}_{p,k}$ and $y_{p,k}$.\footnote{More precisely, $\phivect_{zy}(p,k)$ is composed of $y$ PSD `cross-terms', i.e., $y$ taken at frame $p$ and previous frames, and of $x,y$ cross-PSD terms for $y$ taken at frame $p$ and $x$ taken at previous frames.} 
In practice, these auto- and cross-PSD terms can be estimated by averaging the corresponding auto- and cross-STFT spectra over $D$ frames:
 \begin{align}\label{hphi}
\hat{\phi}_{yy}(p,k) = \frac{1}{D}\sum_{d=0}^{D-1}y_{p-d,k} \: y_{p-d,k}^*.
 \end{align}
The elements in $\phivect_{zy}(p,k)$ can be estimated by using the same principle. Consequently, in practice (\ref{phi}) is approximated as
 \begin{align}\label{hatphi}
 \hat{\phi}_{yy}(p,k) = \hat{\phivect}_{zy}\tp(p,k) \: \mathbf{g}_k.
 \end{align}
Let $P$ denote the total number of the STFT frames. $Q_k$ is the minimum index of $p$ to guarantee that the elements in $\mathbf{z}_{p,k}$ are available from the STFT coefficients of the binaural signals.
For PSD estimation, the previous $D-1$ frames of the current frame are utilized as shown in (\ref{hphi}). Therefore, $p_f=Q_k+D-1$  is the minimum index of $p$ to guarantee that all the frames 
for computing $\hat{\phivect}_{zy}(p,k)$ are available from the STFT coefficients of the binaural signals. 
By concatenating the frames from $p_f$ to $P$, (\ref{hatphi}) can be written in matrix-vector form:
 \begin{align}\label{Phi}
 \hat{\phivect}_{yy}(k) = \hat{\Phimat}_{zy}(k) \: \mathbf{g}_k,
 \end{align}
with
\begin{align}
 &\hat{\phivect}_{yy}(k)=[\hat{\phi}_{yy}(p_f,k),\dots,\hat{\phi}_{yy}(p,k),\dots,\hat{\phi}_{yy}(P,k)]\tp, \nonumber \\ 
 &\hat{\Phimat}_{zy}(k)=[\hat{\phivect}_{zy}(p_f,k),\dots,\hat{\phivect}_{zy}(p,k),\dots,\hat{\phivect}_{zy}(P,k)]\tp. \nonumber 
\end{align}
Note that $\hat{\phivect}_{yy}(k)$ is a $(P-p_f+1)\times1$ vector and $\hat{\Phimat}_{zy}(k)$ is a $(P-p_f+1)\times(2Q_k-1)$ matrix.
In principle, an estimate $\hat{{\mathbf{g}}}_k$ of ${\mathbf{g}}_k$ can be found be solving this linear equation. However, in practice, the sensor signals contain noise and thus the estimated PSD contain noise power. Therefore, we have to remove this noise power before estimating ${\mathbf{g}}_k$.

\section{DP-RTF Estimation in the Presence of Noise}
\label{sec:dprtfn}
Noise always exists in real-world configurations.
In the presence of noise, some frames in (\ref{Phi}) are dominated by noise. Besides, the PSD estimate of speech signals is deteriorated by noise. In this section, an inter-frame subtraction technique enabling to improve the DP-RTF estimation in noise is described, based on a speech frame selection process. 

\subsection{Noisy Signals and PSD Estimates}\label{sec:dprtfn:noise}

In the presence of additive noise (\ref{xn}) becomes
\begin{align}\label{xnu}
\begin{array}{l}
 \tilde{x}(n)=x(n)+u(n)=a(n)\star s(n)+u(n), \\
 \tilde{y}(n)=y(n)+v(n)=b(n)\star s(n)+v(n), 
 \end{array}
\end{align}
where $u(n)$ and $v(n)$, the noise signals, are assumed to be \addnote[exp_noise_stationarity]{1}{individually wide-sense stationary (WSS) and uncorrelated with $s(n)$. Moreover, $u(n)$ and $v(n)$ are assumed to be either uncorrelated, or correlated but jointly WSS. } 
Applying the STFT to the binaural signals in (\ref{xnu}) leads to 
\begin{align*}
\tilde{x}_{p,k} &= x_{p,k}+u_{p,k} \\
\tilde{y}_{p,k} &= y_{p,k}+v_{p,k},
\end{align*}
in which each quantity is the STFT coefficient of its corresponding time-domain signal.
Similarly to ${\mathbf{z}}_{p,k}$, we define
\begin{align} 
 \tilde{\mathbf{z}}_{p,k} &= [\tilde{x}_{p,k},\dots,\tilde{x}_{p-Q_k+1,k},\tilde{y}_{p-1,k},\dots,\tilde{y}_{p-Q_k+1,k}]\tp \nonumber \\
 & =\mathbf{z}_{p,k}+\mathbf{w}_{p,k} \nonumber
\end{align}
where 
\begin{align}
 \mathbf{w}_{p,k}=[u_{p,k},\dots,u_{p-Q_k+1,k},v_{p-1,k},\dots,v_{p-Q_k+1,k}]\tp. \nonumber
\end{align}
The PSD of $\tilde{y}_{p,k}$ is $\phi_{\tilde{y}\tilde{y}}(p,k)$. We define the PSD vector $\phivect_{\tilde{z}\tilde{y}}(p,k)$ composed of the auto- and cross-PSDs between the elements of $\tilde{\mathbf{z}}_{p,k}$ and $\tilde{y}_{p,k}$.
Following (\ref{hphi}), these PSDs can be estimated as $\hat{\phi}_{\tilde{y}\tilde{y}}(p,k)$ and $\hat{\phivect}_{\tilde{z}\tilde{y}}(p,k)$ by averaging the auto- and cross-STFT spectra of input signals over $D$ frames.
Since the speech and noise signals are uncorrelated, we can write
\begin{align}\label{hatphin}
\begin{array}{l}
 \hat{\phi}_{\tilde{y}\tilde{y}}(p,k) = \hat{\phi}_{yy}(p,k)+\hat{\phi}_{vv}(p,k), \\
 \hat{\phivect}_{\tilde{z}\tilde{y}}(p,k) = \hat{\phivect}_{zy}(p,k)+\hat{\phivect}_{wv}(p,k),
 \end{array}
\end{align}
where $\hat{\phi}_{vv}(p,k)$ is an estimation of the PSD of $v_{p,k}$, and $\hat{\phivect}_{wv}(p,k)$ is a vector composed of the estimated auto- or cross- PSDs between the entries of ${\mathbf{w}}_{p,k}$ and ${v}_{p,k}$. 

\subsection{Inter-Frame Spectral Subtraction}\label{sec:dprtfn:ss}

From (\ref{hatphi}) and (\ref{hatphin}), we have for any frame $p$:
\begin{equation}
 \hat{\phi}_{\tilde{y}\tilde{y}}(p,k) - \hat{\phi}_{vv}(p,k) = (\hat{\phivect}_{\tilde{z}\tilde{y}}(p,k) - \hat{\phivect}_{wv}(p,k))\tp \mathbf{g}_k, 
\end{equation}
or alternately:
\begin{equation} \label{linear_noisy}
 \hat{\phi}_{\tilde{y}\tilde{y}}(p,k) = \hat{\phivect}_{\tilde{z}\tilde{y}}(p,k)\tp \mathbf{g}_k  + \hat{\phi}_{vv}(p,k) - \hat{\phivect}_{wv}(p,k)\tp \mathbf{g}_k.\end{equation}
By subtracting the estimated PSD $\hat{\phi}_{\tilde{y}\tilde{y}}(p,k)$ of one frame, e.g. $p_2$, from the estimated PSD of another frame, e.g. $p_1$, we obtain
\begin{align} \label{sub1}
\hat{\phi}_{\tilde{y}\tilde{y}}^{s}(p_1,k) &\triangleq \hat{\phi}_{\tilde{y}\tilde{y}}(p_1,k)-\hat{\phi}_{\tilde{y}\tilde{y}}(p_2,k) \nonumber \\
&= \hat{\phi}_{yy}^s(p_1,k)+e_{vv}(p_1,k)
\end{align}
with
\begin{align*}
 \hat{\phi}_{yy}^s(p_1,k) &= \hat{\phi}_{yy}(p_1,k)-\hat{\phi}_{yy}(p_2,k), \\
 e_{vv}(p_1,k) &= \hat{\phi}_{vv}(p_1,k)-\hat{\phi}_{vv}(p_2,k).
\end{align*}
Applying the same principle to $\hat{\phivect}_{\tilde{z}\tilde{y}}(p,k)$, we have:
\begin{align} \label{sub2}
\hat{\phivect}_{\tilde{z}\tilde{y}}^s(p_1,k)& \triangleq \hat{\phivect}_{\tilde{z}\tilde{y}}(p_1,k)-\hat{\phivect}_{\tilde{z}\tilde{y}}(p_2,k) \nonumber \\
&= \hat{\phivect}_{zy}^s(p_1,k)+\mathbf{e}_{wv}(p_1,k),
\end{align}
with
\begin{align*}
\hat{\phivect}_{zy}^s(p_1,k)&= \hat{\phivect}_{zy}(p_1,k)-\hat{\phivect}_{zy}(p_2,k),  \\
 \mathbf{e}_{wv}(p_1,k) &= \hat{\phivect}_{wv}(p_1,k)-\hat{\phivect}_{wv}(p_2,k). 
\end{align*}
Applying (\ref{linear_noisy}) to frames $p_1$ and $p_2$ and subtracting the resulting equations, we obtain:
\begin{align}\label{hatphis}
\hat{\phi}_{\tilde{y}\tilde{y}}^{s}(p_1,k) &= \hat{\phivect}_{\tilde{z}\tilde{y}}^{s}(p_1,k)\tp \mathbf{g}_k+e(p_1,k),
\end{align}
where 
\begin{align}
e(p_1,k)=e_{vv}(p_1,k)-\mathbf{e}_{wv}(p_1,k)\tp \mathbf{g}_k.
\end{align}
Because $v(n)$ is stationary, $e_{vv}(p_1,k)$ is small. Conversely, the fluctuations of speech signals are much larger than the fluctuations of the noise signal because the speech signals are both non-stationarity and sparse, i.e., speech power spectrum can vary significantly over frames. 
Thence, by properly choosing the frame indexes $p_1$ and $p_2$, for instance in such a way that the speech power $\hat{\phi}_{yy}(p_1,k)$ is high and the speech power $\hat{\phi}_{yy}(p_2,k)$ is low, we have $\hat{\phi}_{yy}^s(p_1,k)\gg e_{vv}(p_1,k)$, or equivalently $\hat{\phi}_{\tilde{y}\tilde{y}}^s(p_1,k)\gg e_{vv}(p_1,k)$. \addnote[exp_jointly]{1}{The same reasoning applies to $\mathbf{e}_{wv}(p_1,k)$, except that the $u$-$v$ cross-terms of $\mathbf{e}_{wv}(p_1,k)$ are small compared to $\hat{\phi}_{\tilde{y}\tilde{y}}^{s}(p_1,k)$ either if $u$ and $v$ are uncorrelated, or if $u$ and $v$ are jointly WSS, which are our (quite reasonable) working assumptions. }

The choice of the frame index necessitates to classify the frames into two sets, $\mathcal{P}_1$ and $\mathcal{P}_2$,
which have high speech power and very low speech power, respectively. This is done in Subsection~\ref{sec:dprtfn:fc} using the minimum and maximum statistics of noise spectrum. 
Before that, we finalize the estimation of the DP-RTF in the noisy case, based on (\ref{hatphis}).

\subsection{DP-RTF Estimation}\label{sec:dprtfn:extraction}

Let $P_1=|\mathcal{P}_1|$ denote the cardinality of $\mathcal{P}_1$. The PSD subtractions (\ref{sub1}) and (\ref{sub2}) are applied to all the frames $p_1\in\mathcal{P}_1$ using their corresponding frames $p_2 \in\mathcal{P}_2$, denoted as $p_2(p_1)$. In practice, $p_2(p_1)$ is the frame in $\mathcal{P}_2$ that is nearest to $p_1$, since the closer the two frames, the smaller the difference of their noise PSD and the difference of their transfer function. 
The resulting PSDs and cross-PSD vectors are gathered into a $P_1\times1$ vector and a $P_1\times(2Q_k-1)$ matrix, respectively, as:
\begin{align}
 \hat{\phivect}_{\tilde{y}\tilde{y}}^s(k)&=[\hat{\phi}_{\tilde{y}\tilde{y}}^{s}(1,k),\dots,\hat{\phi}_{\tilde{y}\tilde{y}}^{s}(p_1,k),\dots,\hat{\phi}_{\tilde{y}\tilde{y}}^{s}(P_1,k)]\tp, \nonumber \\
 \hat{\Phimat}_{\tilde{z}\tilde{y}}^s(k)&=[\hat{\phivect}_{\tilde{z}\tilde{y}}^s(1,k),\dots,\hat{\phivect}_{\tilde{z}\tilde{y}}^s(p_1,k),\dots,\hat{\phivect}_{\tilde{z}\tilde{y}}^s(P_1,k)]\tp. \nonumber
\end{align}
Let us denote $\mathbf{e}(k) = [e(1,k),\dots,e(p_1,k),\dots,e(P_1,k)]\tp$ the $P_1\times1$ vector that concatenates the residual noise for the $P_1$ frames. 
Then, from (\ref{hatphis}) we obtain the following linear equation, which is the ``noisy version'' of (\ref{Phi}):
 \begin{align}\label{Phin}
 \hat{\phivect}_{\tilde{y}\tilde{y}}^s(k) = \hat{\Phimat}_{\tilde{z}\tilde{y}}^s(k)\mathbf{g}_k+\mathbf{e}(k).
 \end{align}
\addnote[exp_noise]{1}{Assuming that the sequence of residual noise entries in $\mathbf{e}(k)$ is i.i.d.\footnote{This assumption is made to simplify the analysis. In practice, $e(p_1,k)$ may be a correlated sequence because of the possible correlation of $\hat{\phi}_{vv}(p,k)$ (or $\hat{\phivect}_{wv}(p,k)$) across frames. Taking this correlation into account would lead to a weighted least square solution to (\ref{Phin}), involving a weight matrix in (\ref{eq:wls}). This weight matrix is not easy to estimate, and in practice, (\ref{eq:wls}) delivers a good estimate of $\hat{g}_{0,k}$, as assessed in our experiments.} 
and also assuming $P_1 \geq (2Q_k-1)$, the least square solution to (\ref{Phin}) is given by:
\begin{align}\label{eq:wls}
 \hat{\mathbf{g}}_k = (\hat{\Phimat}_{\tilde{z}\tilde{y}}^s(k)^H\hat{\Phimat}_{\tilde{z}\tilde{y}}^s(k))^{-1}\hat{\Phimat}_{\tilde{z}\tilde{y}}^s(k)^H\hat{\phivect}_{\tilde{y}\tilde{y}}^s(k),
\end{align}}
where $^H$ denotes matrix conjugate transpose. Finally, the estimation of the DP-RTF $d_k$ defined in \eqref{eq:dp-rtf} is provided by the first element of $\hat{\mathbf{g}}_k$, denoted as $\hat{g}_{0,k}$. 

Note that if two frames in $\mathcal{P}_1$ are close to each other, their corresponding elements in vector $\hat{\phivect}_{\tilde{y}\tilde{y}}^s(k)$ (or corresponding rows in matrix $\hat{\Phimat}_{\tilde{z}\tilde{y}}^s(k)$) will be correlated.
This correlation yields some redundancy of the linear equations. However, in practice, we keep this redundancy to make full use of data and give a more robust solution to (\ref{Phin}).

Still assuming that $e(p_1,k)$ is i.i.d and denoting its variance by $\sigma_k^2$, the covariance matrix of $\hat{\mathbf{g}}_k$ is given by \cite{manolakis2005}:
\begin{align}
 \mathbf{cov}\{\hat{\mathbf{g}}_k\}=\sigma_k^2(\hat{\Phimat}_{\tilde{z}\tilde{y}}^s(k)^H\hat{\Phimat}_{\tilde{z}\tilde{y}}^s(k))^{-1}.
\end{align}
The statistical analysis of the auto- and cross-PSD estimates show that $\sigma_k^2$ is inversely proportional to the number of smoothing frames $D$ \cite{manolakis2005}. 
Thence using a large $D$ leads to a small error variance $\sigma_k^2$. However, increasing $D$ decreases the fluctuation of the estimated speech PSD among frames and thus makes the elements in the matrix $\hat{\Phimat}_{\tilde{z}\tilde{y}}^s(k)^H\hat{\Phimat}_{\tilde{z}\tilde{y}}^s(k)$ smaller, which results in a larger variance of $\hat{\mathbf{g}}_k$. 
Therefore, an appropriate value of $D$ should be chosen to achieve a good tradeoff between smoothing the noise spectrum and preserving the fluctuation of speech spectrum.

\addnote[exp_g]{1}{Finally, to improve the robustness of the DP-RTF estimation, we also calculate (\ref{eq:wls}) after exchanging the roles of the two channels in the whole process. This delivers an estimate $\hat{g}_{0,k}'$ of the inverse of \eqref{eq:dp-rtf}, i.e. an estimate of the inverse DP-RFT $\frac{a_{0,k}}{b_{0,k}}$. Both $\hat{g}_{0,k}$ and ${\hat{g}_{0,k}'}{\inverse}$ are estimates of $\frac{b_{0,k}}{a_{0,k}}$. The final DP-RTF estimate is given by averaging these two estimates as:} 
\begin{equation}
\label{eq:dp-estimation}
\hat{c}_k = \frac{1}{2}(\hat{g}_{0,k}+ {\hat{g}_{0,k}'}{\inverse}).
\end{equation}

\subsection{Frame Classification}\label{sec:dprtfn:fc} 

We adopt the minimum-maximum statistics for frame classification, which was first introduced in \cite{mine2015assp}, and is applied to a different feature in this paper. 
Frame classification is based on the estimation of $\tilde{y}$ PSD, i.e., $\hat{\phi}_{\tilde{y}\tilde{y}}(p,k)$. The frame $p_1$ is selected such that $\hat{\phi}_{\tilde{y}\tilde{y}}^s(p_1,k)$ in (\ref{hatphis}) is large compared to $e(p_1,k)$, and thus (\ref{hatphis}) matches well the noise-free case.

As shown in (\ref{hatphin}), the PSD estimation $\hat{\phi}_{\tilde{y}\tilde{y}}(p,k)$ is composed of both speech and noise powers. 
A minimum statistics formulation was proposed in \cite{martin2001}, where the minimum value of the smoothed periodograms with respect to the index $p$, multiplied by a bias correction factor, is used as the estimation of the noise PSD. 
Here we introduce an equivalent sequence length for analyzing the minimum and maximum statistics of noise spectra, and propose to use two classification thresholds (for two classes $\mathcal{P}_1$ and $\mathcal{P}_2$) defined from the ratios between the maximum and minimum statistics. 
In short, we classify the frames by using the minimum controlled maximum border. 

Formally, the noise power in $\hat{\phi}_{\tilde{y}\tilde{y}}(p,k)$ is 
 \begin{align}\label{hphiv}
\xi_{p,k} \triangleq \hat{\phi}_{vv}(p,k) = \frac{1}{D}\sum_{d=0}^{D-1}|v_{p-d,k}|^2.
 \end{align}
For a stationary Gaussian signal, the probability density function (PDF) of periodogram $|v_{p,k}|^2$ obeys the exponential distribution \cite{martin2001}
\begin{align}
 f(|v_{p,k}|^2;\lambda)=\frac{1}{\lambda}e^{-|v_{p,k}|^2/\lambda}
\end{align}
where $\lambda=E\{|v_{p,k}|^2\}$ is the noise PSD. Assume that the sequence of $|v_{p,k}|^2$ values at different frames are i.i.d. random variables. The averaged periodogram $\xi_{p,k}$ obeys the Erlang distribution \cite{forbes2011statistical} with scale parameter $\mu=\lambda/D$ and shape parameter $D$:
\begin{equation}\label{erl}
f(\xi_{p,k};D,\mu)=\frac{\xi_{p,k}^{D-1}e^{-\frac{\xi_{p,k}}{\mu}}}{\mu^D(D-1)!}. 
\end{equation} 
We are interested in characterizing and estimating the ratio between the maximum and minimum statistics of the sequence $\xi_{p,k}$. Since the maximum and minimum statistics are both linearly proportional to $\mu$ \cite{martin2001}, we assume, without loss of generality, that $\mu=1$. Consequently the mean value of $\xi_{p,k}$ is equal to $D$. 

\begin{figure}[t]
\centering
\vspace{-0.0cm}
\includegraphics[width=0.45\textwidth]{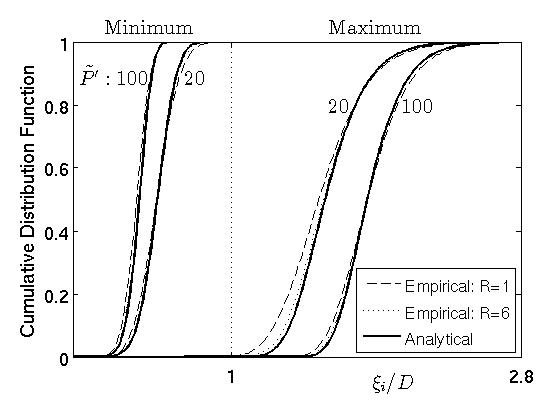}
\caption{\small{Cumulative distribution function (CDF) of the minimum and maximum statistics of $\xi_{p,k}$ for $D=12$.}} 
\label{figmcmt}
\end{figure}

As mentioned in Section \ref{sec:dprtf:estimation}, the frame index of the estimated PSDs $\hat{\phi}_{yy}(p,k)$ and $\xi_{p,k}$ is confined to the range $p_f$ to $P$. 
Let $R$ denote the increment of the frame index $p$ of the estimated PSDs. 
If $R$ is equal to or larger than $D$, for two adjacent estimated PSD $\xi_{p,k}$ and $\xi_{p+R,k}$, there is no frame overlap. The sequence $\xi_{p,k}, \ p=p_f:R:P$ is then an independent random sequence. 
The length of this sequence is $\tilde{P}=\lceil\frac{P-p_f+1}{R}\rceil$. 
The PDFs of the minimum and maximum of these $\tilde{P}$ independent variables are \cite{martin1994}:
\begin{equation}\label{fmin}
\begin{array}{l}
f_{min}(\xi) = \tilde{P}\cdot(1-F(\xi))^{\tilde{P}-1}\cdot f(\xi),  \\ 
f_{max}(\xi) = \tilde{P}\cdot F(\xi)^{\tilde{P}-1} \cdot f(\xi),
\end{array}
\end{equation} 
where $F(\cdot)$ denotes the cumulative distribution function (CDF) associated with the PDF (\ref{erl}).
Conversely, if $R<D$, $\xi_{p,k}$ is a correlated sequence, and the correlation coefficient is linearly proportional to the frame overlap.
\addnote[exp_e37]{1}{For this case, (\ref{fmin}) will not be valid anymore. Based on a large amount of simulations using white Gaussian noise (WGN),\footnote{The simulations are done with the following procedure: applying STFT to a number of WGN signals with identical long duration. 
For each time-frequency bin, estimate the PSD by averaging the periodograms of the past $D$ frames. 
Without loss of generality, the scale parameter $\mu$ of the PSD estimation can be set to 1 by adjusting the noise PSD $\lambda$ to $D$.
A sequence of correlated PSD estimates is generated by picking PSD estimates from the complete sequence, with frame increment $R$ (with $R < D$). The length of the correlated sequence is $\tilde{P}$.
The minimum/maximum values of each correlated sequence are collected at each frequency for all the WGN signals. The PDF and CDF of the minimum/maximum statistics are simulated by the histograms of these minimum/maximum values. 
Fig. \ref{figmcmt} shows some examples of this empirical CDF.} 
it was found that the following approximate equivalent sequence length
\begin{equation}\label{f12}
\tilde{P}' = \frac{\tilde{P}R}{D}\cdot\left(1+\textrm{log}\left(\frac{D}{R}\right)\right)
\end{equation}
can replace $\tilde{P}$ in order to make (\ref{fmin}) valid for the correlated sequence.
We observe that the ratio between the number $D$ of frames used for spectrum averaging and the frame increment $R$ of PSD estimates, is replaced with its logarithm. Note that this is an empirical result, for which theoretical foundation remains to be investigated.}

Then, the expectation of the minimum can be approximately computed as
\begin{equation}\label{f13}
\bar{\xi}_{min} \approx \frac{\sum\nolimits_{\xi_i}\xi_i\cdot f_{min}(\xi_i)}{\sum\nolimits_{\xi_i}f_{min}(\xi_i)},
\end{equation}
where $\xi_i\in\{0,0.1D,0.2D,\dots,3D\}$ is a grid used to approximate the integral operation, which well covers the support of the Erlang distribution with shape $D$ and scale 1. 
Similarly, the CDF of the maximum can be estimated as 
\begin{align}
F_{max}(\xi) \approx \sum\nolimits_{\xi_i} f_{max}(\xi_i).
\end{align}
Finally, we define two classification thresholds that are two specific values of the maximum and minimum ratios, namely 
\begin{align}
r_1=\frac{\xi_{F_{max}(\xi)=0.95}}{\bar{\xi}_{min}}, \mbox{ and } r_2=\frac{\xi_{F_{max}(\xi)=0.5}}{\bar{\xi}_{min}}, 
\end{align}
where $\xi_{F_{max}(\xi)=0.95}$ and $\xi_{F_{max}(\xi)=0.5}$ are the values of $\xi$ for which the CDF of the maximum is equal to 0.95 and 0.5, respectively.
Classes $\mathcal{P}_1$ and $\mathcal{P}_2$ are then obtained with 
\begin{align}\label{f15}
\mathcal{P}_1 &= \{ p \ | \ \xi_{p,k}> r_1 \cdot \min_p\{\xi_{p,k}\} \}, \\
\mathcal{P}_2 &= \{ p \ | \ \xi_{p,k}\le r_2 \cdot \min_p\{\xi_{p,k}\} \}.
\end{align}
These two thresholds are set to ensure that the frames in $\mathcal{P}_1$ contain large speech power and the frames in $\mathcal{P}_2$ 
contain negligible speech power. The speech power for the other frames are probabilistically uncertain, making them unsuitable for either $\mathcal{P}_1$ or $\mathcal{P}_2$. Using two different thresholds evidently separates speech region and noise-only region. In other words, there is a low probability to have a frame classified into $\mathcal{P}_1$ in the proximity of $\mathcal{P}_2$ frames, and vice versa.
Therefore, in general, the PSD of a frame in $\mathcal{P}_1$ is estimated using $D$ frames that are not included in the noise-only region, and vice versa. 
Note that if there are no frames with speech content, e.g., during long speech pauses, class $\mathcal{P}_1$ will be empty with a probability of 0.95 due to threshold $r_1$. 

As an illustration of (\ref{f12}), Fig. \ref{figmcmt} shows the CDF for $D=12$. The empirical curves are simulated using WGN, and the analytical curves are computed using the equivalent sequence length in (\ref{f12}). The minimum CDF and maximum CDF of two groups of simulations are shown, for which the equivalent sequence lengths $\tilde{P}'$ are fixed at 20 and 100, respectively. 
For each equivalent sequence length $\tilde{P}'$, two empirical curves with frame increment $R=1$ and $R=6$ are simulated using WGN, whose corresponding original sequence lengths are $\tilde{P}=69$ and $\tilde{P}=24$ for $\tilde{P}'=20$, and $\tilde{P}=344$ and $\tilde{P}=118$ for $\tilde{P}'=100$, respectively. 
This shows that the equivalent sequence length in (\ref{f12}) is accurate for the minimum and maximum statistics.

\section{Sound Source Localization method}
\label{sec:ssl}
The amplitude and the phase of DP-RTF represent the amplitude ratio and phase difference between two source-to-microphone direct-path ATFs. In other words, in case of two microphones, the DP-RTF is equivalent to the interaural cues, ILD and IPD, associated to the direct path. More generally, we consider here $J$ microphones. This is a slight generalization that will directly exploit the previous developments, since we consider these $J$ microphones pair-wise.
As in \cite{araki2007,mine2015eusipco}, we consider the normalized version of the DP-RTF estimate \eqref{eq:dp-estimation} between microphones $i$ and $j$:
\begin{align}\label{bfc}
 c_{k,ij}=\frac{\hat{c}_{k,ij}}{\sqrt{1+|\hat{c}_{k,ij}|^2}}.
\end{align}
Compared to the amplitude ratio, the normalized DP-RTF is more robust. In particular, when the reference transfer function $a_{0,k}$  is much smaller than $b_{0,k}$, 
the amplitude ratio estimation is sensitive to noise present in the reference channel. By concatenating  \eqref{bfc} across $K$ frequencies and across $(J-1)J/2$ microphone pairs, we obtain a high-dimensional feature vector $\mathbf{c}\in\mathbb{R}^{J(J-1)K/2}$. 
Since speech signals have a sparse STFT representation, we denote by $\hvect\in\mathbb{C}^{J(J-1)K/2}$ an indicator vector whose elements are either equal to 1 if the energy at the corresponding frequency is significant, or equal to 0 if the energy is negligible. 
\addnote[exp_under]{1}{In practice, the indicator vector entries at a given frequency $k$ are set to 0 if the corresponding matrix $\hat{\Phimat}_{\tilde{z}\tilde{y}}^s(k)$ is underdetermined, i.e. $P_1<(2Q_k-1)$ for that frequency. This way, we do not use any DP-RTF calculated from (\ref{eq:wls}) for such ``missing frequency'' (see below).}

\addnote[cla_doa]{1}{The proposed DP-RTF estimation method is suitable for the most general case of microphone setup where the microphones are not necessarily placed in free-field. In other words it can be applied to any microphone pair in any microphone array setup. For instance, in the present paper, the microphones are placed in the ears of a dummy head or on the head of a robot. In these cases, there is no clear (analytical) relationship between the HRIR/HRTF/DP-RTF and the DOA of the emitting source, even after removal of the noise and reverberations. } In order to perform SSL based on the feature vector $\mathbf{c}$, we adopt here a supervised framework: A training set 
$D_{\mathbf{c,q}}$ of $I$ pairs $\{\mathbf{c}_i, \mathbf{q}_i\}_{i=1}^I$ is available, where $\mathbf{c}_i$ is a DP-RTF feature vector 
generated with an anechoic head-related impulse response (HRIR),
and $\mathbf{q}_i$ is the corresponding source-direction vector. 
Then, for an observed (test) feature vector $\mathbf{c}$ that is extracted from the microphone signals, the corresponding direction is estimated using either (i)~nearest-neighbor search in the training set (considered as a look-up table) or (ii)~a regression whose parameters have been tuned from the training set. 
\addnote[cla_tt]{1}{Note that the training set and the observed test features should be recorded using the same microphone set-up. This way, the HRIR of the training set (corresponding to an anechoic condition) corresponds to the direct-path of the BRIR of the test condition (recorded in reverberant condition). } 

Nearest-neighbor search corresponds to solving the following minimization problem (\addnote[cla_odot]{1}{$\odot$ denotes the Hadamard product, i.e. entry-wise product}):
\begin{equation}\label{f16}
\hat{\mathbf{q}} = 
 \mathop{\textrm{argmin}}_{i \in [1,I]} \parallel \hvect \odot ( \mathbf{c}-\mathbf{c}_i)  \parallel.  
\end{equation}
\addnote[exp_under2]{1}{As mentioned above, the indicator vector $\hvect$ enables to select the relevant DP-RTF vector components, i.e. the ones corresponding to frequencies with (over)determined solution to (\ref{Phin}).  }
Because of the sparse nature of the test feature vectors, not any regression technique could be used. Indeed, one needs a regression method that allows training with full-spectrum signals and testing with sparse-spectrum signals. Moreover, the input DP-RTF vectors are high dimensional and not any regression method can handle high-dimensional input data. For these reasons we adopted the probabilistic piece-wise linear regression technique of 
\cite{deleforge2015acoustic}.

\section{Experiments with Simulated Data}
\label{sec:experiments1}
We report results with experiments carried out in order to evaluate the performance of the proposed method. We simulated various experimental conditions in terms of reverberation and additive noise.

\subsection{The Dataset}
\begin{figure}[t!]
\centering
\includegraphics[width=0.40\textwidth]{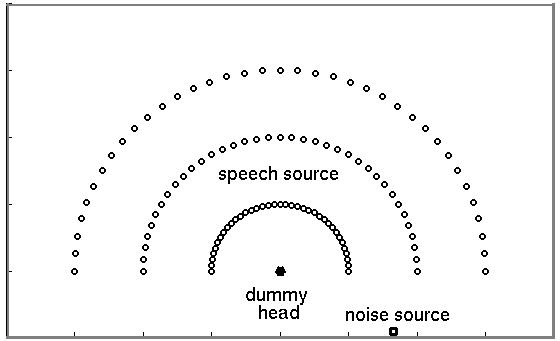}
\caption{\small{Configurations of room, dummy head, speech sources and noise source for the BRIR dataset.}} 
\label{figroom}
\end{figure}

The BRIRs are generated with the ROOMSIM simulator \cite{campbell2004} and with the head related transfer function (HRTF) of a KEMAR dummy head \cite{gardner1995}. 
The responses are simulated in a rectangular room of dimension $8$~m~$\times$~$5$~m~$\times$~$3$~m. The KEMAR dummy head is located at $(4, 1, 1.5)$~m. The sound sources are placed in front of the dummy head with 
azimuths varying from $-90^\circ$ to $90^\circ$, spaced by 5$^\circ$, an elevation of 0$^\circ$, and distances of $1$~m, $2$~m, and $3$~m., see Fig.\ref{figroom}.

The absorption coefficients of the six walls are equal, and adjusted to control $T_{60}$ at 0.22~s, 0.5~s and 0.79~s, respectively. Two other quantities, i.e. the ITDG and the direct-to-reverberation ratio (DRR), are also important to
measure the intensity of the reverberation. In general, the larger the source-to-sensors distance is, the smaller the ITDG and DRR are. For example, when $T_{60}$ is 0.5~s, the DRRs for $1$, $2$, $3$~m are about $1.6$, $-4.5$ and $-8.1$ dB, respectively.
Speech signals from the TIMIT dataset \cite{garofolo1988} are used as the speech source signals, which are convolved with the simulated BRIRs to generate the sensor signals.
Each BRIR is convolved with 10 different speech signals from TIMIT to achieve reliable SSL results. 
Note that the elevation of the speech sources is always equal to $0^\circ$ in the BRIR dataset, thence in these simulated-data experiments the source direction corresponds to the azimuth only.
The feature vectors in the training set $\{\mathbf{c}_i\}_{i=1}^I$ are generated with the anechoic HRIRs of the KEMAR dummy head from the azimuth range $[-90^\circ\, , 90^\circ]$, spaced by 5$^\circ$, i.e. $I=37$. 
In this section, the nearest-neighbor search is adopted for localization.

Two types of noise signals are generated: (i)~a ``directional noise'' is obtained by convolving a single channel WGN signal with a BRIR corresponding to position beside the wall with azimuth of $120^\circ$, elevation of $30^\circ$ and distance of $2.2$~m, see Fig. \ref{figroom}; (ii)~an ``uncorrelated noise'' consists of an independent WGN signal on each channel.  Noise signals are added to the speech sensor signals with various signal-to-noise ratios. 

\begin{table*}[t]
\centering
\caption{\small{Localization errors (degrees) for different values of $Q$ in different conditions. $T_{60}=0.5$~s. ``Distance'' stands for source-to-sensors distance.  The bold value is the minimum localization error for each condition.}}
\begin{tabular}{| ccc |ccccccc|}
\hline
\multicolumn{3}{|c|}{Conditions}         & \multicolumn{7}{c|}{$Q/T_{60}$ ($T_{60}=0.5$ s)}     \\ 
Noise type & SNR & Distance                              & 0.1       & 0.15     & 0.2      & 0.25    & 0.3    & 0.35   & 0.4            \\  \hline
Uncorrelated & $10$~dB& 1 m	& 0.122      & 0.081    & \textbf{0.077}     & 0.081  & 0.099  &  0.108  & 0.113          \\ 
Uncorrelated & $10$~dB& 2 m  	& 1.338      & 0.847    & 0.716    & 0.649   & 0.629  & 0.608  & \textbf{0.568}        \\ 
Directional& $10$~dB& 1 m    & 0.135      & \textbf{0.113}      & 0.122   & 0.131  & 0.149  &  0.158  &  0.162          \\ 
Directional& $10$~dB& 2 m     & 1.437      & 0.869    & 0.829    & 0.680   & 0.644  & 0.626  & \textbf{0.617}   \\ 
Uncorrelated& $-5$~dB& 2 m    & 7.824      &6.833     & 6.703    &\textbf{6.680}    & 6.802  &6.964   & 7.149      \\ 
Directional& $-5$~dB& 2 m     & 13.36      &12.25     & 11.90    & 11.23   & 10.96  & 10.52  &\textbf{10.38}     \\ \hline
\end{tabular}
\vspace{0mm}
\label{tQ}
\end{table*}

\begin{table*}
\centering
\caption{\small{Localization errors (degrees) for different values of $D$ in different conditions. $T_{60}=0.5$~s. ``Distance'' stands for source-to-sensors distance. The bold value is the minimum localization error for each condition.}}
\begin{tabular}{| c cc | c  c  c  c  cccc|}
\hline
\multicolumn{3}{|c|}{Conditions}         & \multicolumn{8}{c|}{$D$ frames}     \\ 
Noise type & SNR & Distance                              & 6        & 8     & 10      & 12    & 14    & 16   & 18   & 20         \\  \hline
Uncorrelated & $-5$ dB&  1 m	& 2.59   & 2.15  & 2.09   & 1.99   &  1.86  &  1.81  & 1.64   & \textbf{1.59}        \\ 
Uncorrelated & $-5$ dB&  2 m  & 7.37   & 6.03  & 6.17   & 6.68   &  6.08  &  6.40 & 6.90    & 6.50        \\ 
Directional& $-5$ dB&   1 m  & 3.83  &  3.42  &  3.51  &  3.23  &  3.70  &  3.47  &  2.96 & 3.45        \\ 
Directional& $-5$ dB&  2 m   &\textbf{9.80} &  10.28 &  10.32 &  11.23 &  11.60  & 13.18  & 13.62 &15.35  \\ \hline
\end{tabular}
\vspace{0mm}
\label{tD}
\end{table*}

\subsection{Setting the Parameters}
\label{sec:experiments1:parameter}

The sampling rate is $16$~kHz. Only the frequency band from $0$ to $4$~kHz is considered for speech source localization. 
\addnote[settings]{1}{The setting of all three parameters $N$, $Q_k$ and $D$ is crucial for a good estimation of the DP-RTF. 
Intuitively, $Q_k$ should correspond to the value of $T_{60}$ at the $k$-th frequency bin. For simplicity, we set $Q_k$ to be the same for all frequencies and denote it as $Q$. In the following of this subsection, we present preliminary SSL experiments that were done in order to tune $N$, $Q$ and $D$ to an ``optimal tradeoff'' setting that would ensure good SSL performance for a large range of acoustic conditions. 
Since considering all possible joint settings of these three parameters is a hard task, when exploring the setting of one of them, we may fix the others. }

In all the following, the localization error is taken as the performance metric. It is computed by averaging the absolute errors between the localized directions and their corresponding ground truth (in degrees) over the complete test dataset.

Let us first consider the setting of $Q$. Here we fix $N=256$ with $50\%$ overlap, and $D=12$. Table \ref{tQ} shows the localization errors for $Q$ values corresponding to CTF length $\in [0.1T_{60}, \dots, 0.4T_{60}]$ with $T_{60}=0.5$~s. 
When the SNR is high (first four lines; SNR~=~$10$~dB), the influence of noise is small, and the DRR plays a dominant role. 
Comparing the localization errors for source-to-sensors distances between $1$~m and $2$~m, we see that small localization errors are obtained with rather small $Q$ values for $1$ m, and with the larger $Q$ values for $2$~m. This result indicates that, for a given $T_{60}$, $Q$ should be increased when the DRR is decreased. The CTF should cover most of the energy of the room impulse response. By comparing the results for the uncorrelated noise of 10~dB and $-5$~dB, source at 2~m (second and fifth lines), 
we observe that the smallest localization error is achieved by a smaller $Q$ for the low SNR case, compared to the high SNR case. 
Note that a larger $Q$ corresponds to a greater model complexity, which needs more reliable (less noisy) data to be estimated. 
The intense uncorrelated noise degrades the data, thence a small $Q$ is preferred. In contrast, for the directional noise, a large $Q$ is also suitable for the low SNR case (sixth line). The reason is possibly that the
directional noise signal has a similar convolution structure as the speech signal, and the noise residual $\mathbf{e}(k)$ also has a similar convolution structure. Thence the data reliability is not degraded much. 
In conclusion, the optimal $Q$ varies with the $T_{60}$, DRR, noise characteristics, and noise intensity. In practice, it is difficult to obtain these features automatically, thence we assume that $T_{60}$ is known, and we set $Q$ to correspond to $0.25T_{60}$ as a tradeoff for different acoustic conditions. 

Let us now consider the setting of $D$. Here, we set  $Q$ to correspond to $0.25T_{60}$, and $N=256$ with $50\%$ overlap. The number of frames $D$ is crucial for an efficient spectral subtraction (Section \ref{sec:dprtfn:ss}). A large $D$ yields a small noise residual. However, the remaining speech power after spectral subtraction 
may also be small because of the small fluctuations of the speech PSD estimate between frames when $D$ is large. Table \ref{tD} shows the localization errors for $D\in [6, \dots, 20]$ under different conditions. Note that only the results for the low SNR case ($-5$~dB)
are shown, for which the effect of noise suppression plays a more important role. It can be seen (first line) that a large $D$ yields the smallest localization error, which means that removing noise power is more important than retaining speech power for this condition.
The reason is that the DRR is large for source-to-sensors distance of $1$~m, so that the direct-path speech power is relatively large. As $D$ increases, the remaining direct-path speech power decreases only slightly, compared to the decrease of the noise residual.  
In contrast, a small $D$ yields the smallest localization error for the directional noise at $2$~m (fourth line), which means that retaining speech power is more important than removing noise power for this condition.
The reasons are that (i)~as described above, the data reliability is not degraded much by the directional noise in the sense of convolution, and (ii)~the direct-path speech power is relatively small for a source-to-sensors distance of $2$~m.
The conditions of the second and third lines fall in between the first line and the fourth line, and these results do not strongly depend on $D$. 
It is difficult to choose a $D$ value that is optimal for all the acoustic conditions. In the following, we set $D=12$ frames ($100$~ms) as a fair tradeoff. 

As for the setting of $N$, let us remind that the reflections present in $a(n)|_{n=0}^N$ lead to a biased definition of DP-RTF. In order to minimize the reflections contained in $a(n)|_{n=0}^N$, the STFT window length $N$ should be as small as possible, while still capturing the direct-path response. 
However, in practice, a small $N$ requires a large $Q$ for the CTF to cover well the room impulse response, which increases the complexity of the DP-RTF estimate.
We tested the localization performance for three STFT window sizes: $8$~ms ($N=128$ samples), $16$~ms ($N=256$ samples), and $32$~ms ($N=512$ samples), with $50\%$ overlap. Again, $Q$ corresponds to $0.25T_{60}$. For example, with $T_{60}=0.79$~s and with $N=128$, $256$, $512$ respectively, $Q$ is equal to $50$, $25$, $13$ frames respectively. $D$ is set to $100$~ms. For $N=128$, $256$, $512$, $D$ is $24$, $12$, $6$ frames, respectively. Table \ref{tN} shows the localization errors under various acoustic conditions. 
We first discuss the case of high SNR (first three lines). When the source-to-sensors distance is small ($1$~m; first line), the ITDG is relatively large and we observe that $N=128$ and $N=256$ ($8$~ms and $16$~ms windows) achieve comparable performance.
This indicates that, if the ITDG is relatively large, there are not much more reflections in $a(n)|_{n=0}^N$ for a $16$-ms window, compared with an $8$-ms window.
The next results (second line) show that, when $T_{60}$ is small ($0.22$~s), the localization performance decreases much more for a $16$-ms and a $32$-ms window than for an $8$-ms window, as the sensor-to-noise distance increases from $1$~m to $3$~m. A lower ITDG yields a larger DP-RTF estimation error due to the presence of more reflections in $a(n)|_{n=0}^N$. 
When $T_{60}$ increases to $0.79$~s, $Q$ becomes larger, especially for $N=128$. It can be seen (third line) that here $N=256$ yields a better performance than other values. This is because the lack of data leads to a large DP-RTF estimation error for $N=128$, and the reflections in $a(n)|_{n=0}^N$ bring a large DP-RTF estimation error for $N=512$. When the SNR is low ($-5$ dB; last three lines), less reliable data are available due to noise contamination. In that case, a large $N$ achieves the best performance. 
Finally, we set $N=256$ ($16$-ms STFT window) as a good overall tradeoff between all tested conditions.

\setlength{\tabcolsep}{5.0pt}
\begin{table}[t!]
\centering
\caption{\small{Localization errors (degrees) for three values of $N$. ``Distance'' is the sensors-to-source distance. The bold value is the minimum localization error. In this experiment, the noise signal is generated by summing the directional noise and uncorrelated noise with identical powers.} }
\label{tN}
\begin{tabular}{| c c  c| c  c  c  |}
\hline
 \multicolumn{3}{|c|}{Conditions}          & \multicolumn{3}{c|}{STFT window length $N$}     \\ 
 SNR    & Distance  & $T_{60}$     & 128 (8 ms) & 256 (16 ms) & 512 (32 ms)      \\  \hline
        & $1$~m & $0.22$~s 	&\textbf{0.01}    &\textbf{0.01}   & 0.02         \\ 
$10$ dB   & $3$~m& $0.22$ s   &\textbf{0.58}    &1.19   & 1.89         \\ 
        & $3$~m &$0.79$ s	&9.60    &\textbf{9.22}   & 9.55        \\ \hline
        & $1$~m &$0.22$ s	&1.89    &1.62   & \textbf{1.49}  \\ 
$-5$ dB    & $3$~m &$0.22$ s	&8.07    &\textbf{6.30}  & 7.04 \\
        & $3$~m &$0.79$ s	&22.66   &20.81  & \textbf{17.75} \\ \hline  
\end{tabular}
\end{table}

\subsection{DP-RTF Estimation}

\begin{figure*}[t!]
\centering
\includegraphics[width=0.3\textwidth]{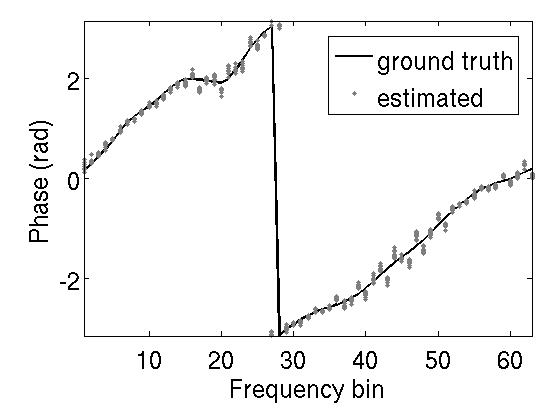}
\includegraphics[width=0.3\textwidth]{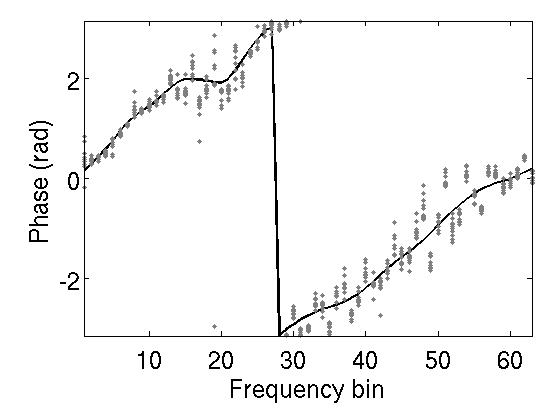}
\includegraphics[width=0.3\textwidth]{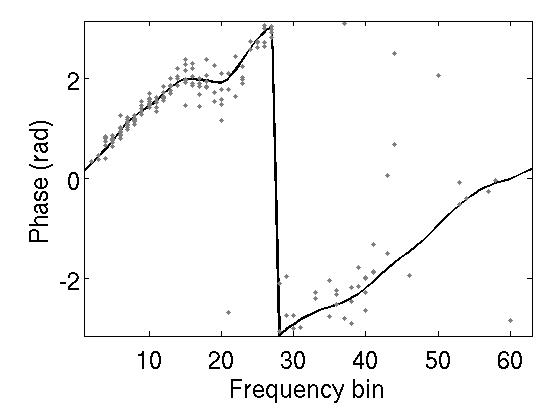} \\
\includegraphics[width=0.3\textwidth]{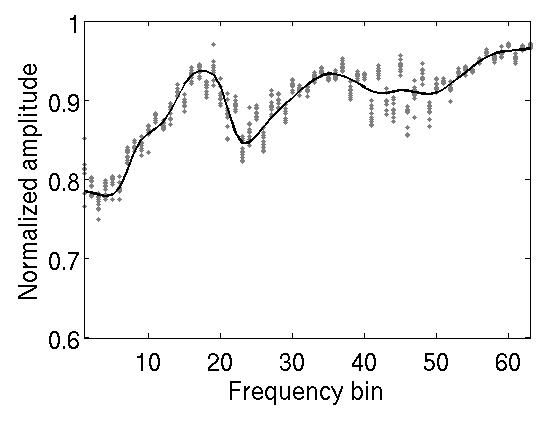}
\includegraphics[width=0.3\textwidth]{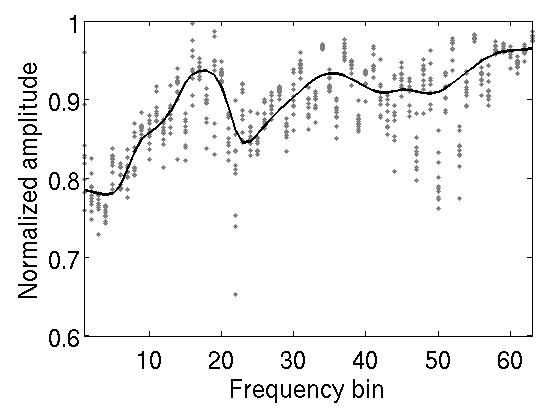}
\includegraphics[width=0.3\textwidth]{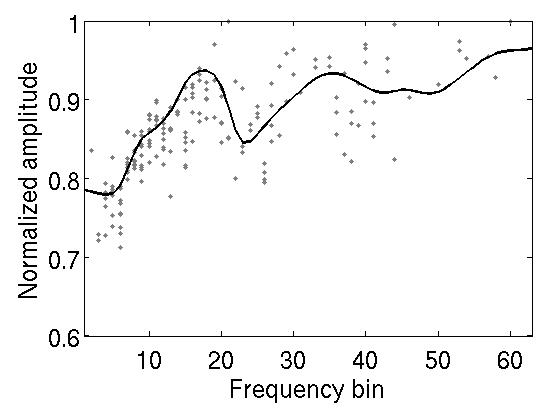}
\caption{\small{The phase (top) and normalized amplitude (bottom) of the normalized estimated DP-RTF (\ref{bfc}) as a function of frequency bins. 
 The source direction is 30$^\circ$. $T_{60}=0.5$ s. The continuous curve corresponds to the ground-truth DP-RTF $d_k$ computed from the anechoic HRTF. 
Left: $1$~m source-to-sensors distance, $30$~dB SNR. Middle: $2$~m source-to-sensors distance, $30$~dB SNR. Right: $1$~m source-to-sensors distance, $0$~dB SNR.
For each acoustic condition, the BRIR  is convolved with 10 different speech recordings as the sensor signals, whose DP-RTF estimations are all shown.
In this experiment, the noise signal is generated by summing the directional noise and uncorrelated noise with identical powers.}} 
\label{figdprtf}
\end{figure*}

We provide several representative examples showing the influence of both reverberation and noise on the DP-RTF estimates. 
The phase and normalized amplitude of the estimated DP-RTF for three acoustic conditions are shown in Fig. \ref{figdprtf}. The SNR is set to $30$~dB in the first two examples, hence the noise is negligible. 
The difference between the estimated and the ground-truth phase is referred to as the phase estimation error. It can be seen that, for most frequency bins, the mean value (over ten trials) of the phase estimation error is very small (but nonzero, which  
indicates that the estimated DP-RTF is biased). As mentioned above, the bias is brought in by the reflections in the impulse response segment $a(n)|_{n=0}^N$. 
In addition, if the DRR gets smaller, a longer CTF is required to cover the room impulse response. However, for a given $T_{60}$, the CTF length $Q$ is set as a constant, for instance $0.25 T_{60}$. In this example, this improper value of $Q$ leads to an inaccurate CTF model, which causes the DP-RTF estimate bias. 
When the source-to-sensors distance increases, both the ITDG and DRR become smaller. Therefore, for both phase and amplitude, the estimation bias of the second example of Fig. \ref{figdprtf} (middle) is larger than the bias of the first example (left).
Moreover, the DP-RTF $\frac{b_{0,k}}{a_{0,k}}$ in $\mathbf{g}_k$ plays a less important role relative to other elements, with decreasing DRR, 
which makes the variance of both the phase and amplitude estimation errors to be larger than in the first example.
By comparing the first and last examples of Fig. \ref{figdprtf}, it is not surprising to observe that the estimation error increases as noise power increases. 
When the SNR is low, less reliable speech frames are available in the high frequency band, due to the intense noise. Therefore, there is no DP-RTF estimation for the frequency bins satisfying $P_1<2Q_k-1$.

\subsection{Baseline Methods}

\addnote[cla_exp1]{1}{In our previous work \cite{mine2015assp}, the proposed inter-frame spectral subtraction scheme was applied to RTF estimators (as opposed to the DP-RTF estimators proposed in the present paper). The results were compared with the RTF estimators proposed in \cite{gannot2001} and \cite{cohen2004} in the presence of WGN or babble noise. The efficiency of the inter-frame spectral subtraction to remove the noise was demonstrated. Thence, the focus of the present set of experiments is mainly aimed at
(i) comparing the robustness to reverberation of the proposed DP-RTF feature with respect to other features, in a similar SSL framework, and at (ii) comparing the proposed SSL method with a conventional SSL method.}   

To this aim, we compare our method with three other methods: (i) an unbiased RTF identification method \cite{mine2015assp}, in which  
a spectral subtraction procedure (similar to the one described in Section~\ref{sec:dprtfn:ss}) is used to suppress noise. Since this RTF estimator is based on the MTF approximation, we refer to this method as RTF-MTF.
(ii) a method based on a STFT-domain coherence test (CT) \cite{mohan2008}.\footnote{Note that \cite{faller2004} introduces a similar technique based on interaural coherence, using features extracted from band-pass filter banks. 
Also, a binaural coherent-to-diffuse ratio approach was proposed in \cite{schwarz2015coherent,zheng2015} and applied to dereverberation but not to SSL.} We refer to this method as RTF-CT. The coherence test is used in \cite{mohan2008} to search the rank-1 time-frequency bins which are supposed to be dominated by one active source. 
We adopt the coherence test for single speaker localization, in which one active source denotes the direct-path source signal. The TF bins that involve notable reflections have low coherence.
We first detect the maximum coherence over all the frames at each frequency bin, and then set the coherence test threshold for each frequency bin to $0.9$ times its maximum coherence. In our experiments, this threshold achieves the best performance.
The covariance matrix is estimated by taking a $120$~ms ($15$ adjacent frames) averaging. \addnote[exp_ctss]{1}{The auto- and cross-PSD spectral subtraction is applied to the frames that have high speech power and a coherence larger than the threshold, and then are averaged over frames for RTF estimation.}  \addnote[exp_srp]{1}{ (iii) a classic one-stage algorithm: the steered-response power (SRP) utilizing the phase transform (PHAT) \cite{dibiase2001,do2007}. The azimuth directions $-90^\circ:5^\circ:90^\circ$ are taken as the steering directions, and their HRIRs are used as the steering responses. }
 
 \addnote[cla_mc]{1}{Note that for both RTF-MTF and RTF-CT methods, the features used in the SSL are obtained after the inter-frame spectral subtraction procedure. The SSL method presented in Section \ref{sec:ssl} is adopted. The training set used as a look-up table or used for training the regression is the same as for the DP-RTF.}

\subsection{Localization Results}

\begin{figure}[!t]
\centering
\includegraphics[width=0.3\textwidth]{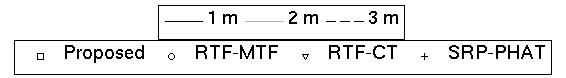}\\
\includegraphics[width=0.45\textwidth]{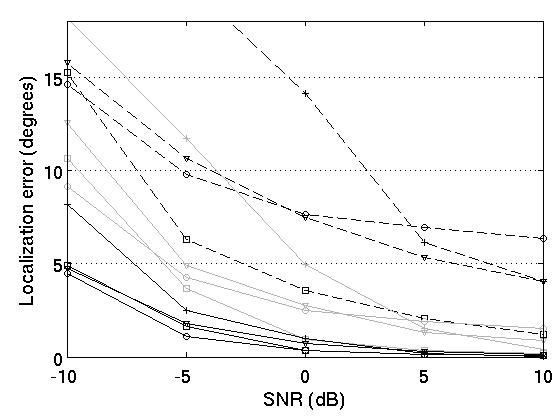}\\
\includegraphics[width=0.45\textwidth]{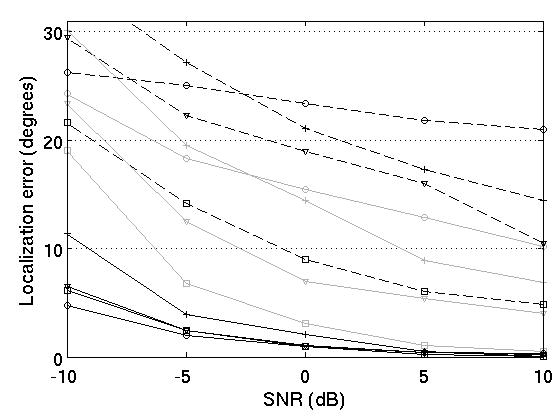}\\
\includegraphics[width=0.45\textwidth]{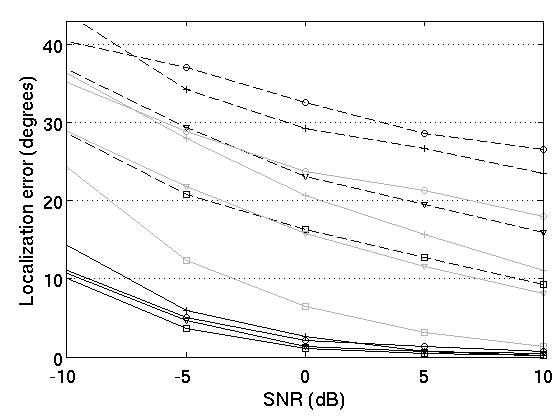}
\caption{\small{Localization errors under various reverberation and noise conditions. Top: $T_{60}=0.22$ s. Middle: $T_{60}=0.5$ s. Bottom: $T_{60}=0.79$ s. 
The localization errors are shown as a function of SNR for source-to-sensors distances of $1$~m, $2$~m  and $3$~m.}}  
\label{fig:simulated-results}
\end{figure}

Fig. \ref{fig:simulated-results} shows the localization results in terms of localization error (let us remind that this error is an average absolute error between the localized directions and their corresponding ground truth (in degrees) over the complete test dataset). Note that in real world, directional noise source, e.g. fan, refrigerator, etc., and diffuse background noise co-exist. Thence in this experiment, the noise signal was generated by summing the directional noise and uncorrelated noise with identical powers.

Let us first discuss the localization performance shown in Fig. \ref{fig:simulated-results}-top for $T_{60}=0.22$ s. When the DRR is high ($1$~m source-to-sensors distance; solid-line), compared with the proposed method, RTF-MTF has a comparable performance under high SNR conditions, and a slightly better performance under low SNR conditions (lower than $0$~dB). 
This indicates that when the reverberation is low, the MTF approximation is valid. When less reliable data are available (under low SNR conditions), the proposed method perform slightly worse than RTF-MTF due to its greater model complexity. Note that both the RTF-MTF and the proposed DP-RTF methods achieve very good localization performance: The localization error goes from almost $0^\circ$ at SNR = $10$~dB to about $5^\circ$ at SNR = $-10$~dB.
RTF-CT achieves the worst performance. This indicates that when the direct-path impulse response is slightly contaminated by the reflections, employing all the data (as done by RTF-MTF and DP-RTF) obtains a smaller localization error than employing only the data selected by the coherence test.  
In general, for mild reverberations, the performance gap between RTF-MTF, RTF-CT and the proposed method is small and the noise level plays a decisive role for good localization.  

\addnote[exp_phat1]{1}{The SRP-PHAT method achieves comparable performance measures with the three other methods when the SNR is high ($10$~dB). However, the performance measures of SRP-PHAT degrades immediately and dramatically when the SNR decreases. 
The steered-response power is severely influenced by intense noise, especially by the directional noise. This indicates that the inter-frame spectral subtraction algorithm applied to RTF-MTF, RTF-CT and the proposed method is efficient to reduce the noise.  }

When the DRR decreases ($2$~m source-to-sensors distance, grey lines; $3$~m source-to-sensors distance, dashed lines), the performance measures of RTF-MTF degrades notably. 
For SNR~=~$10$~dB, the localization error of RTF-MTF increases from $0.07^\circ$ to $1.51^\circ$ and to $6.35^\circ$ for source-to-sensors distances of $1$~m, $2$~m and $3$~m, respectively. 
The direct-path impulse response is severely contaminated by the reflections. 
At high SNRs, RTF-CT performs slightly better than RTF-MTF. Indeed, RTF-CT selects the frames that contain less reverberations for calculating the RTF estimate, which improves the performance at high SNR conditions. However, when the noise level increases, the precision of RTF-CT also degrades. The performance of RTF-CT is influenced not only by the residual noise but also by the decline of the coherence test precision, which make it fall even faster than RTF-MTF with decreasing SNR (it has a larger localization error at $-5$ dB and $-10$ dB).    

The proposed method also has a larger localization error when the source-to-sensors distance increases:
the DP-RTF estimation is possibly influenced by the increased amount of early reflections in the impulse response segment $a(n)|_{n=0}^N$, by the effect of an improper $Q$ setting, and by the decreased importance of $\frac{b_{0,k}}{a_{0,k}}$ in vector $\mathbf{g}_k$. 
However, the performance of the proposed DP-RTF method degrades much slower than the ones of RTF-MTF when the source distance increases. 
For an SNR of $10$~dB, the localization error of the proposed method increases from $0.06^\circ$ to $0.16^\circ$ and $1.19^\circ$ as the source-to-sensors distance increases from $1$~m to $2$~m and $3$~m.
It can be seen that the performance of the proposed method also falls faster than RTF-MTF with decreasing SNR, since the available data is less reliable.
The localization error of the proposed method is larger than the MTF error at -10 dB.
It is observed that the proposed method prominently outperforms RTF-CT. It is shown in \cite{nadiri2014} that the coherence test is influenced by the coherent reflections (very early reflections) of the source signal. Moreover, it is difficult to automatically set a coherence test threshold that could perfectly select the desired frames.
Many frames that have a coherence larger than the threshold include reflections.

\addnote[exp_phat2]{1}{The performance of SRP-PHAT also degrades with the DRR decrease. It is known that PHAT-based method are quite sensible to reverberations and noise in general. Briefly, the performance measures of SRP-PHAT are in between the performance measures of RTF-MTF and RTF-CT for high SNRs, which indicates that the PHAT weight could suppress the reverberations only to a certain extent. Below $5$~dB, SRP-PHAT performs worst of the four methods. }

Fig. \ref{fig:simulated-results}~(bottom) displays the results for $T_{60}=0.79$ s. Obviously, the performance measures of all four methods degrade as $T_{60}$ increases. Indeed, the MTF approximation is not accurate; there are only a few time-frequency bins with a rank-1 coherence; and a large value of $Q$ has to be utilized in the proposed method, for which there may not always be enough reliable data. 
Here, it can be seen that RTF-CT performs better than RTF-MTF for any SNR value and source-to-sensors distance. \addnote[exp_phat3]{1}{Even SRP-PHAT performs better than RTF-MTF (for $2$~m and $3$~m source-to-sensors distance). } This shows that the RTF estimation error brought by the MTF approximation largely increases as $T_{60}$ increases. For $1$~m source-to-sensors distance, the proposed method performs slightly better than all other three methods. For $2$~m and $3$~m source-to-sensors distance, the proposed method largely outperforms the other three methods, at all SNRs. For example, at SNR~=~$0$~dB, the proposed method achieves about $6.5^\circ$ of localization error at $2$~m source-to-sensors distance, while RTF-CT (the best of the three baseline methods) achieves about $15.8^\circ$, hence the gain for the proposed method over the best baseline is about $9.3^\circ$.  
However, the performance of the proposed method and of RTF-CT still have a faster degradation with decreasing SNR compared to RTF-MTF. 

Finally, we can see from Fig. \ref{fig:simulated-results}~(middle), that the performance of the different methods for $T_{60}=0.5$~s falls in between the other two cases shown on the same figure, and the trends of performance evolution with $T_{60}$ is consistent with our comments above.

In summary, the proposed method outperforms the three other methods under most acoustic conditions. In a general manner, the gain over the baseline methods increases as the source-to-sensors distance increases (or the DRR decreases) and as the reverberation time increases (but the influence of the noise level is more intricate). As a result, the proposed method achieves acceptable localization performance in quite adverse conditions. 
For example (among many others), with $T_{60}=0.5$~s, source-to-sensors distance of $3$~m and an SNR of $0$~dB, the localization error is about $9^\circ$, and with $T_{60}=0.79$~s, source-to-sensors distance of $2$~m, and an SNR of $0$~dB, the localization error is about $6.5^\circ$.

In all the above results, the duration of the signal used for localization was not considered with great attention: The localization errors were averaged over 10 sentences of TIMIT of possibly quite different duration, from $1$~s to $5$~s. Yet the number of available frames that are used to construct (\ref{Phin}) depends on the speech duration, which is crucial for the least square DP-RTF estimation in (\ref{eq:wls}). Here we complete the simulation results with a basic test of the influence of the speech duration on localization performance. To this aim we classified our TIMIT test sentences according to their duration (closer to $1$~s, $2$~s, $3$~s or $4$~s) and proceeded to localization evaluation for each new group (of 10 sentences), for a limited set of acoustic conditions (SNR~=~$10$~dB and $0$~dB, $T_{60}=0.5$~s).  
Table \ref{td} shows the localization errors of the proposed method, the RTF-MTF, and the RTF-CT method, for the four tested approximate speech durations. We can see that, as expected, all three methods achieve a smaller localization error when increasing speech duration, for both tested SNRs. 
The improvement is more pronounced for the proposed method and the RTF-CT method compared to the RTF-MTF method. For example, for SNR~=~$10$~dB, the localization error is reduced by $66\%$ (from $1.57^\circ$ to $0.54^\circ$) for the proposed method, and by $49\%$ (from $6.24^\circ$ to $3.21^\circ$) for the RTF-CT method when the speech duration rises from $1$~s to $4$~s. In contrast, the localization error of RTF-MTF is quite larger and is only reduced by $11\%$ (from $12.60^\circ$ to $11.16^\circ$).

\begin{table}[t!]
\caption{\small{Localization errors (in degrees) as a function of speech duration, for $T_{60}=0.5$~s and a source-to-sensors distance of $2$~m.}}
\label{td}
\centering
\begin{tabular}{| c  c| c  c  c  c  |}
\hline
   &      & \multicolumn{4}{c|}{Speech duration (s)}     \\ 
 SNR & Method      & 1 & 2 & 3 & 4       \\  \hline
        &Proposed	&1.57    &0.88   & 0.79  &   0.54        \\ 
$10$ dB   & RTF-CT     &6.24    &4.43   & 3.86  &   3.21       \\ 
        & RTF-MTF	&12.60   &12.01  & 11.25 &  11.16       \\ \hline
        & Proposed	&7.36    &4.62   & 4.05  &  3.07  \\ 
$0$ dB    & RTF-CT	&12.97   &11.33  & 10.04 &   9.67 \\
        & RTF-MTF	&17.56   &15.29  & 14.94 &  15.01 \\ \hline  
\end{tabular}
\end{table}

\section{Experiments with the NAO Robot}
\label{sec:experiments2}

\begin{figure}[t]
\centering
{\includegraphics[width=0.8\columnwidth]{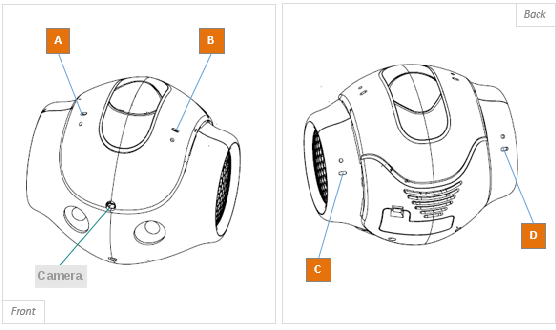}}
\caption{\small{NAO's head has four microphones and one camera.}} 
\label{fignao}
\end{figure}

In this section we present several experiments that were conducted using the NAO robot (Version 5) in various real-world environments. NAO is a humanoid companion robot developed and commercialized by Aldebaran Robotics.\footnote{https://www.ald.softbankrobotics.com.} NAO's head has four microphones that are nearly coplanar, see Fig.~\ref{fignao}.  
The recordings contain ego-noise, i.e. noise produced by the robot. In particular, it contains a loud fan noise, which is stationary and partially interchannel correlated \cite{loellmann2014}. The spectral energy of the fan noise is notable up to 4 kHz, thence the speech signals are significantly contaminated. Note that the experiments reported in this section adopt the parameter settings discussed in Section \ref{sec:experiments1:parameter}. 

\subsection{The Datasets}
\begin{figure}[t]
\centering
\includegraphics[height=0.31\columnwidth]{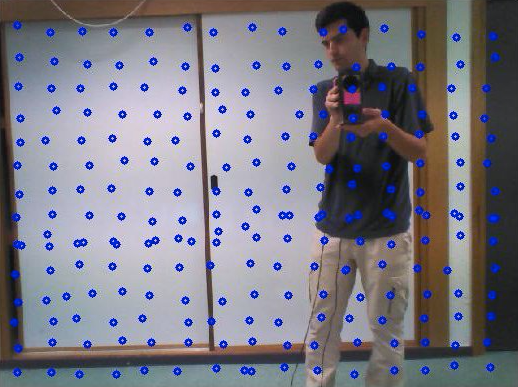}
\hspace{0.05cm}
\includegraphics[height=0.31\columnwidth]{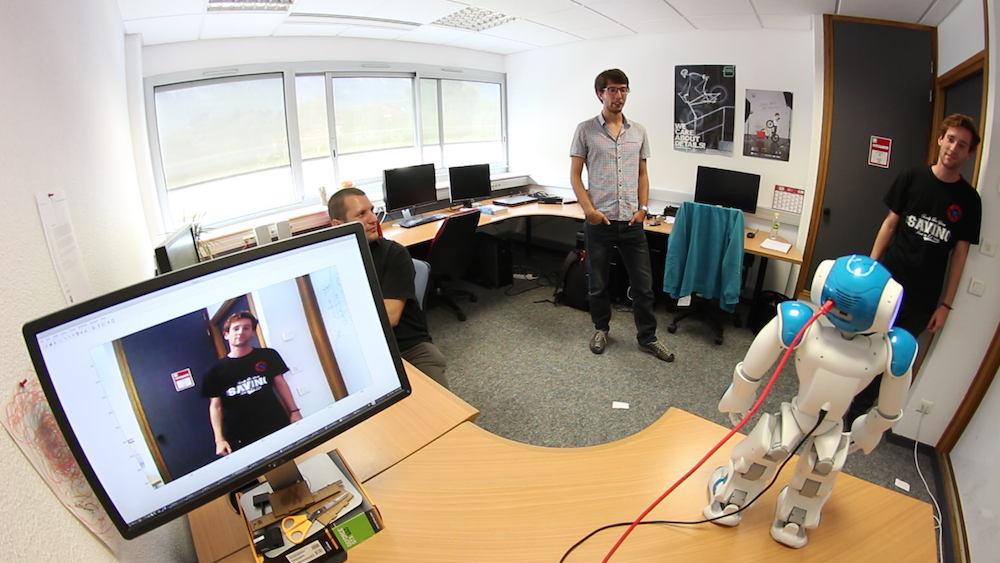}
\caption{ \small{
The \textit{audio-visual} training dataset (left) is obtained by moving a loudspeaker in front of a microphone/camera setup. Sounds are emitted by a loudspeaker. A LED placed on the loudspeaker enables to associate each sound direction with an image location (a blue circle). The data contain pairs of acoustic recordings and sound directions.
A typical localization scenario with the NAO robot (right). 
}} 
\label{fig:scenario}
\end{figure}

The data are recorded in three environments: laboratory, office, e.g., Fig.~\ref{fig:scenario}-(right), and cafeteria, 
with reverberation times ($T_{60}$) that are approximately $0.52$~s, $0.47$~s and $0.24$~s, respectively.
Two \textbf{test datasets} are recorded in these environments: \\
1) The \emph{audio-only} dataset: In the laboratory, speech utterances from the TIMIT dataset \cite{garofolo1988} are emitted by a loudspeaker in front of NAO. 
Two groups of data are recorded with a source-to-robot distance of $1.1$~m and $2.1$~m, respectively. 
For each group, $174$ sounds are emitted from directions uniformly distributed in azimuth and elevation, in the range $[-120^\circ,120^\circ]$ (azimuth), and $[-15^\circ, 25^\circ]$ (elevation).\\
2) The \emph{audio-visual} dataset: Sounds are emitted by a loudspeaker lying in the field of view of NAO's camera.
The image resolution is of $640 \times 480$ pixels, corresponding to approximately $60^\circ$ ($-30^\circ$ to $30^\circ$) azimuth range and to approximately $48^\circ$ ($-24^\circ$ to $24^\circ$) elevation range, so $1^\circ$ of azimuth/elevation corresponds to approximately $10.5$ horizontal/vertical pixels. A LED placed on the loudspeaker enables to estimate the loudspeaker location in the image, hence ground-truth localization data are available with the audio-visual dataset.
Three sets of audio-visual data are recorded in three different rooms. For each set, sounds are emitted from about $230$ directions uniformly distributed in the camera field-of-view.
Fig.~\ref{fig:scenario}-(left) shows the source positions shown as blue dots in the image plane. The source-to-robot distance is about $1.5$~m in this dataset.

In both datasets, ambient noise is much lower than fan noise, hence the noise of recorded signals mainly corresponds to fan noise. 
In the case of the audio-only dataset, the SNR is $14$~dB and $11$~dB for source-to-robot distances of $1.1$~m and $2.1$~m, respectively. For the audio-visual dataset the SNR is $2$~dB.

The \textbf{training dataset} 
for the \emph{audio-only} localization experiments is generated with the NAO head HRIRs of $1,002$ directions uniformly distributed over the same azimuth-elevation range as the test dataset. The training dataset for \emph{audio-visual} experiments is generated with the NAO head HRIR of $378$ directions uniformly distributed over the camera field-of-view. 
HRIRs are measured in the laboratory: white Gaussian noise is emitted from each direction, and the cross-correlation 
between the microphone and source signals yields the BRIR of each direction. In order to obtain anechoic HRIRs, the BRIRs are manually truncated before the first reflection.
The regression method of \cite{deleforge2015acoustic}, outlined in Section~\ref{sec:ssl}, is used for supervised localization.
The SRP-PHAT method takes the source directions in the training set as the steering directions.

\subsection{Localization Results for the \textit{Audio-Only} Dataset}

Experiments with the audio-only dataset first show that elevation estimation in the range $[-15^\circ\, 25^\circ]$ is unreliable for all the four methods. This can be explained by the fact that the four microphones are coplanar.
Therefore we only present the azimuth estimation results in the following.   

\begin{figure}[t]
\centering
\includegraphics[width=0.95\columnwidth]{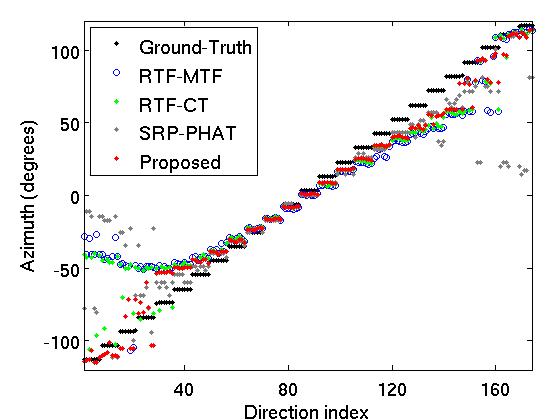} \\
\includegraphics[width=0.95\columnwidth]{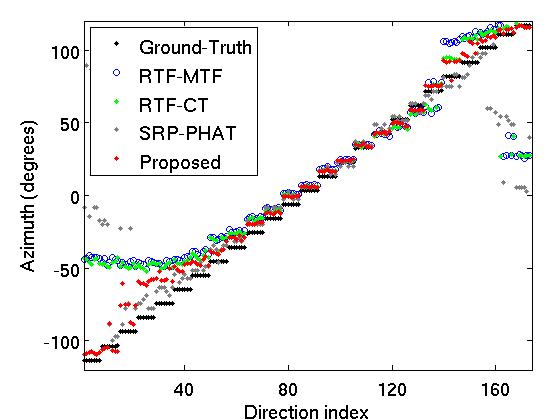}
\caption{\small{Azimuth estimation for the audio-only dataset. Source-to-robot distance is $1.1$~m (top) and $2.1$~m (bottom). 
}} 
\label{figres}
\end{figure}

The azimuth estimation results for the audio-only dataset are given in Fig.~\ref{figres}. The results are quite consistent across the two conditions, i.e. source-to-robot distance of $1.1$~m (Fig.~\ref{figres}-top) and $2.1$~m (Fig.~\ref{figres}-bottom). Globally, for the azimuth range $[-50^\circ,50^\circ]$ all four methods provide good localization, i.e. they follow the ground-truth line quite well, for both source-to-robot distances. In this range, the proposed method achieves slightly better results than the RTF-MTF and RTF-CT methods. The performance of all methods drops significantly for directions out of this range, but globally, the proposed method remains the closest to the ground-truth. \addnote[exp_phat4]{1}{In more details, in the approximate range $[-120^\circ,-50^\circ]$ and $[50^\circ,120^\circ]$ it can be seen that SRP-PHAT and RTF-MTF have the largest localization error and many localization outliers caused by reverberations (SRP-PHAT performs slightly better than RTF-MTF in the zones just after $-50^\circ$ and $50^\circ$, possibly due to PHAT weighting
). } By selecting frames that involve less reverberations, RTF-CT performs slightly better than RTF-MTF. 
The proposed method outperforms the others by extracting the binaural cues associated with the direct-path propagation. Importantly, in the extremities of the range, the proposed method does not generate major outliers nor large deviation from the ground-truth, as opposed to the other methods.

\subsection{Localization Results for the Audio-Visual Dataset}


The azimuth and elevation in the audio-visual dataset are limited to a small range around $0^\circ$ azimuth. As a consequence, both the azimuth and elevation localization results of this dataset are better than the results of audio-only dataset in average.
Table~\ref{tle} shows the localization errors for azimuth (Azim.) and elevation (Elev.) for the audio-visual dataset.  
The elevation errors are always larger than the azimuth errors, due to the low elevation resolution of the microphone array that we already mentioned (the microphone are coplanar and the microphone plane is horizontal). 
The cafeteria has the smaller reverberation time, $T_{60}=0.24$~s. Consequently, the RTF-MTF and RTF-CT methods yields performance measures that are comparable with the proposed method. The office and laboratory have larger reverberation times,  $0.47$~s and $0.52$~s, respectively, so the MTF approximation is no more accurate. A bit surprisingly RTF-MTF performs better than RTF-CT for the office (though the errors are quite close), this is probably  due to the fact that the coherence test does not work well under low SNR conditions (let us remind that the SNR of the audio-visual dataset is around $2$~dB). \addnote[exp_phat5]{1}{Globally, SRP-PHAT performs the worst, due to the intense noise. }
As a result of the presence of notable reverberations, the proposed method performs here significantly better than the three other methods. For example, in the laboratory environment, the proposed method provides  $0.84^\circ$ azimuth error and $1.84^\circ$ elevation error, vs. $1.41^\circ$ azimuth error and $2.30^\circ$ elevation error for the best baseline methods (for instance SRP-PHAT and RTF-MTF respectively).

\begin{table}[t!]
\caption{\small{Localization error (in degrees) for the audio-visual dataset. The best results are shown in bold.}}
\label{tle}
\centering
\begin{tabular}{| c | c  c | c  c | c  c |}
\hline         
	    & \multicolumn{2}{c|}{Cafeteria} & \multicolumn{2}{c|}{Office} & \multicolumn{2}{c|}{Laboratory}   \\
 Method    & Azim.      & Elev.      & Azim.      & Elev.     & Azim.     & Elev.                                   \\ \hline
 RTF-MTF        & 0.47      & 1.58      & 0.62     & 2.14     & 1.46   & 2.30                                      \\
 RTF-CT         & \textbf{0.43}  & 1.49      & 0.68     & 2.30     & 1.59   & 2.40                                      \\ 
 SRP-PHAT        & 0.77      & 1.95      & 1.03     &2.80          & 1.41  & 3.33                                     \\
 Proposed   & 0.48      & \textbf{1.46}      & \textbf{0.55}     & \textbf{1.86}     & \textbf{0.84}   & \textbf{1.84}                                      \\ \hline
\end{tabular}
\end{table}


\section{Conclusion}\label{sec:conclusion}

We proposed a method for the estimation of the direct-path relative transfer function (DP-RTF). Compared with the conventional RTF, the DP-RTF is defined as the ratio between two direct-path acoustic transfer functions. Therefore, the DP-RTF definition and estimation implies the removal of the reverberations, and it provides a more reliable feature, in particular for sound source localization. To estimate the DP-RTF, we adopted the convolutive transfer function (CTF) model instead of the multiplicative transfer function (MTF) approximation. By doing this, the DP-RTF can be estimated by solving a set of linear equations constructed from the reverberant sensor signals. Moreover, an inter-frame spectral subtraction method was proposed to remove noise power. 
This spectral subtraction process does not require explicit estimation of the noise PSD, hence it does not suffer from noise PSD estimation errors. 

Based on the DP-RTF we proposed a supervised sound-source localization algorithm. \addnote[exp_tra]{1}{The latter relies on a training dataset that is composed of pairs of DP-RTF feature vectors and their associated sound directions. The training dataset is pre-processed in such a way that it only contains anechoic head-related impulse responses. Hence the training dataset does not depend on the particular acoustic properties of the recording environment. Only the sensors set-up must be consistent between training and testing (e.g. using the same dummy/robot head). }
In practice we implemented two supervised methods, namely a nearest-neighbor search and a mixture of linear regressions. Experiments with both simulated data and real data recorded with four microphones embedded in a robot head, showed that the proposed method outperforms an MTF-based method and a method based on a coherence test, \addnote[exp_phat6]{1}{as well as a conventional SRP-PHAT method, } in reverberant environments.


In the presented experiments the model parameters $Q$, $D$ and $N$ (Section \ref{sec:experiments1:parameter}) were set to constant values which were chosen as a tradeoff yielding good results in a variety of acoustic conditions. In the future, to improve the robustness of DP-RTF, we plan to estimate
the acoustic conditions using the microphone signals, such that an optimal set of parameters can be adaptively adjusted.  We also plan to extend the DP-RTF estimator and its use in SSL to the more complex case of multiple sound sources.
\end{document}